\documentstyle[aps,twocolumn]{revtex}

\input{epsf}
\newcommand{\eps}{\varepsilon}

\begin{document}

\title{Pole structure of the Hamiltonian $\zeta$-function for a
        singular potential}%
\author{H.\ Falomir$^a$, P.\ A.\ G.\ Pisani$^a$ and A.\ Wipf$^b$ }%
\address{$a)$ IFLP, Departamento de F\'{\i}sica - Facultad de Ciencias Exactas, UNLP
C.C. 67, (1900) La Plata, Argentina \\
$b)$ Theoretisch--Physikalisches Institut,
        Friedrich--Schiller--Universit\"at Jena, Max--Wien--Platz 1,
        07743 Jena, Germany}%

\date{December 11, 2001}%

\maketitle

% ----------------------------------------------------------------
\begin{abstract}

We study the pole structure of the $\zeta$-function associated to
the Hamiltonian $H$ of a quantum mechanical particle living in the
half-line $\mathbf{R}^+$, subject to the singular potential $g
x^{-2}+x^2$. We show that $H$ admits nontrivial self-adjoint
extensions (SAE) in a given range of values of the parameter $g$.
The $\zeta$-functions of these operators present poles which
depend on $g$ and, in general, do not coincide with half an
integer (they can even be irrational). The corresponding residues
depend on the SAE considered.

\bigskip



\end{abstract}
% ----------------------------------------------------------------

\section{Introduction}

In Quantum Field Theory under external conditions, quantities like
effective actions and vacuum energies, which describe the
influence of external fields or boundaries on the physical system,
are generically divergent and require a renormalization to get a
physical meaning.

In this context, a powerful and elegant regularization scheme is
based on the use of spectral functions, such as the associated
$\zeta$-function \cite{Dowker,Hawking} and heat-kernel (for recent
reviews see, for example, \cite{Elizalde,Bytsenko,Klaus,Bordag}).

It is well known \cite{Seeley,Gilkey} that  for an elliptic
boundary value problem in a $\nu$-dimensional compact manifold
with boundary, described by a differential operator $A$ of order
$\omega$, with smooth coefficients and defined on a domain of
functions subject to local boundary conditions, the
$\zeta$-function
\begin{equation}\label{zeta-func-def}
  \zeta_A(s)\equiv Tr\{A^{-s}\}
\end{equation}
has a meromorphic extension to the complex $s$-plane presenting
isolated simple poles at $s=(\nu-j)/\omega$, with $j=0,1,2,\dots$

In the case of positive definite operators, the $\zeta$-function
is related, via Mellin transform, to the trace of the heat-kernel
of the problem, $Tr\{e^{-t A}\}$. The pole structure of
$\zeta_A(s)$ determines the small-$t$ asymptotic expansion of this
trace \cite{Gilkey},
\begin{equation}\label{heat-trace}
  Tr\{e^{-t A}\}\sim \sum_{j=0}^\infty a_j(A)\, t^{(j-\nu)/\omega},
\end{equation}
where the coefficients are related to the residues by
\begin{equation}\label{coef-res}
  a_j(A)=\left.{\rm Res}\right|_{s=(\nu-j)/\omega}
  \Gamma(s)\,\zeta_A(s).
\end{equation}

However, for the case of a differential operator with coefficients
presenting singularities, less is known about the structure of the
$\zeta$-function or the heat-kernel trace asymptotic expansion.

Callias \cite{Callias1,Callias2,Callias3} has argued that, when
the coefficient in the zero-th order term in an elliptic,
(essentially) self-adjoint, second order differential operator
presents a singularity like $1/x^2$, the heat-kernel trace
asymptotic expansion in terms of  powers $t^{(j-\nu)/2}$ (as in
(\ref{heat-trace})) is ill-defined, and an expansion including
$\log t$ and perhaps more general powers of $t$ ($t^\alpha$ with
$\alpha\neq n/2$) would be in order. In particular, considering
Hamiltonians $H$ with these characteristics, it has been given in
\cite{Callias1,Callias2} a small-$t$ asymptotic expansion for the
diagonal element $e^{-t\, H}(x,x)$ which also presents $
t^{(j-\nu)/2}\, \log t$ terms, and where some of the coefficients
are distributions with support concentrated at the singularities.

\bigskip

It is the aim of the present article to analyze the pole structure
of the $\zeta$-function of a Hamiltonian $H$ describing a quantum
Schr\oe dinger particle living in the half-line $\mathbf{R}^+$,
subject to a singular potential given by $V(x)=g \, x^{-2}+ x^2$,
for a real $g$ \cite{foot}.% with $g\in\mathbf{R}$.

For a certain range of values of $g$, this Hamiltonian (a second
order differential operator) admits nontrivial self-adjoint
extensions\footnote{The existence of nontrivial SAE for this kind
of singular potential has been pointed out in \cite{Simon-73}. SAE
with more general singular potentials have also been considered in
\cite{Rellich,Albeverio}.} in $\mathbf{L_2}(\mathbf{R^+})$, each
one describing a different physical system. We will show that the
associated $\zeta$-function presents isolated simple poles which
depend on $g$, which (in general) do not lie at $s=(1-j)/2$ for
$j=0,1,\dots$, and can even be irrational numbers. Moreover, we
will find that the residues at these simple poles depend on the
self-adjoint extension of $H$ considered.

This pole structure for the $\zeta$-function implies a small-$t$
asymptotic expansion for the heat kernel trace of the problem in
terms of powers which (in general) are not half an integer.
Moreover, the coefficients in this expansion depend on the
selected self-adjoint extension.

\bigskip

The structure of the paper is the following: In Section
\ref{adjoint-H} we specify the adjoint of the Hamiltonian operator
and in Section \ref{deficiency-sub} we determine its deficiency
subspaces. The Hamiltonian self-adjoint extensions are
characterized in Section \ref{SAE-H}, and in Section
\ref{spectrum-SAE} is described the corresponding spectrum. In
Section \ref{integ-rep} we give an integral representation for the
$\zeta$-function of each SAE of the Hamiltonian and in Section
\ref{pole-structure} we discuss the structure of its
singularities. In Section \ref{particular} we analyze some
particular cases, and we establish our conclusions in Section
\ref{conclusions}. Appendix \ref{closure} is devoted to the
construction of the closure of $H$, and in Appendix
\ref{asymptotic} we outline the necessary asymptotic expansions.

\section{The Hamiltonian and its adjoint}  \label{adjoint-H}

Let us consider the operator
\begin{equation}\label{ham}
    H=-\frac{d^2}{dx^2}+V(x),
\end{equation}
with
\begin{equation}
    V(x)=\frac{g}{x^2}+x^2,
\end{equation}
densely defined on the domain ${ \mathcal D}(H)={\mathcal
C}_0^\infty(\mathbf{R}^+)$, the linear space of functions
$\varphi(x)$ with continuous derivatives of all order and compact
support non containing the origin. It is easily seen that $H$ is a
symmetric operator.

In order to construct the SAE \cite{Reed-Simon} of $H$ we must get
its adjoint, $H^\dagger$, and determine the deficiency subspaces.

The operator $H^\dagger$ is defined on the subspace of
square-integrable functions $\psi(x)$ for which $(\psi, H
\varphi)$ is a continuous linear functional of $\varphi\in
{\mathcal D}(H)$. This requires the existence of $\chi(x)\in
\mathbf{L_2}(\mathbf{R}^+)$ such that $(\psi, H
\varphi)=(\chi,\varphi), \forall\, \varphi\in {\mathcal D}(H)$.
If this is the case, then $\chi(x)$ is uniquely defined, since
${\mathcal D}(H)$ is dense in $\mathbf{L_2}(\mathbf{R}^+)$ and, by
definition, $H^\dagger \psi = \chi$.

For $\psi\in {\mathcal D}(H^\dagger)$ and $\forall\,
\varphi\in{\mathcal D}(H)$ we have
\begin{equation}\label{distrib}
  \begin{array}{c}
    (\psi,H\varphi) = \int_0^{\infty}\psi(x)^*
    (-\varphi''(x)+V(x)\, \varphi(x))\, dx = \\ \\
    = \left((-\psi''+V(x)\psi),\varphi \right) =
    (\chi,\varphi),
  \end{array}
\end{equation}
where the derivatives of $\psi$ are taken in the sense of
distributions.

Equation (\ref{distrib}) implies that
$\psi''(x)=V(x)\psi(x)-\chi(x)$, is a locally integrable function.
Then, its primitive $\psi'(x)$ is absolutely continuous for $x>0$.

Therefore \cite{Prop.2}, the domain of $H^\dagger$ is the subspace
of square integrable functions having an absolutely continuous
first derivative and such that
\begin{equation}\label{Hmas}
  H^\dagger\psi(x)= -\psi''(x)+V(x) \psi(x)\in
\mathbf{L_2}(\mathbf{R}^+)
\end{equation}
(without requiring any boundary condition at $x=0$).

\bigskip

In  the next Section we will determine the deficiency subspaces
of $H$, ${\mathcal K}_{\pm}= {\rm Ker}(H^\dagger \mp i )$.

\section{Deficiency subspaces of H } \label{deficiency-sub}

To compute the deficiency indices \cite{Reed-Simon} of $H$,
$n_\pm = {\rm dim}\,{\mathcal K}_{\pm}$, we must solve the
eigenvalue problem
\begin{equation}\label{eigequ}
    H^\dagger\phi_{\lambda}=-\phi_\lambda''(x)+
    V(x) \phi_\lambda(x)=\lambda \phi_{\lambda},
\end{equation}
for $\phi_{\lambda}\in {\mathcal D}(H^\dagger)$ and
$\lambda\in\mathbf{C}$, with the imaginary part $\Im(\lambda)\neq
0$.

By means of the following Ansatz (suggested by the expected
behavior of the solutions of (\ref{eigequ}) for $x \rightarrow
0^+$ and $x\rightarrow \infty$),
\begin{equation}
    \phi=x^{\alpha}e^{-\frac{x^2}{2}}F(x^2),
\end{equation}
with
\begin{equation}\label{alpha}
  \alpha=1/2+\sqrt{g+1/4},
\end{equation}
we get from (\ref{eigequ}) the Kummer's equation for F(z):
\begin{equation}\label{Kum}
    zF''(z)+(b-z)F'(z)-aF(z)=0,
\end{equation}
where $a={(2\alpha+1-\lambda)}/{4}$ and $b=\alpha+{1}/{2}$.

For real  $\alpha$, we have  $g\geq -1/4$ and $\alpha\geq 1/2$. In
this case it can be seen \cite{Abramowitz} that the only solution
of eq.\ (\ref{Kum}) leading to a square-integrable at infinity
solution of eq.\ (\ref{eigequ}) is given by the Kummer function
$F(z)=U(a;b;z)$. Then, the eigenfunctions of $H^\dagger$ are
proportional to
\begin{equation}\label{eigfun}
    \phi_{\lambda}(x)=x^{\alpha}\, e^{-\frac{x^2}{2}}\,
    U\left(\frac{2\alpha+1-\lambda}{4};
    \alpha+\frac{1}{2};x^2\right).
\end{equation}

We must now study the behavior of $\phi_{\lambda}$ near the
origin, where  $U(a;b;z)$ behaves  as $z^{-a}(1+ O(1/z))$
\cite{Abramowitz}. We must consider two different regions for the
parameter $\alpha$.

\bigskip

For $\alpha\geq 3/2$, $\phi_{\lambda}\in {\mathbf L_2}({\mathbf
R}^+) \Leftrightarrow  a ={(2\alpha+1-\lambda)}/{4}=-n$, with
$n\in \mathbf{N}$. As a consequence, if $\lambda\notin
\mathbf{R}$, $\phi_{\lambda}\notin L_2(\mathbf{R}^+)$, and the
deficiency subspaces are trivial.

This means that, for $\alpha\geq 3/2$, $H$ is essentially
self-adjoint, and its discrete spectrum is given by the condition
$-a\in\mathbf{N}$, i.\ e.\,
\begin{equation}\label{spec-ESA}
    \lambda_n=4n +2\alpha+1,
\end{equation}
with $n=0,1,2,\ldots$ The corresponding eigenfunctions are
\begin{equation}\label{eigenfunc}
    \phi_n=x^{\alpha}e^{-\frac{x^2}{2}}
    U\left(-n;
    \alpha+\frac{1}{2};x^2\right).
\end{equation}

\bigskip

On the other hand, for $1/2\leq\alpha<3/2$, one can see
\cite{Abramowitz} that $\phi_{\lambda}\in L_2(\mathbf{R}^+),
\forall \lambda\in\mathbf{C}$. Then, the deficiency subspaces
${\mathcal K}_\pm$ are one-dimensional, and the deficiency indices
$n_\pm=1$\footnote{This is in accordance to Weyl's criterion
\cite{Reed-Simon} according to which, for continuous $V(x)$, $H$
is essentially self-adjoint if and only if it is in the limit
point case, both at infinity and at the origin.

In addition, if $V(x)\geq M>0$, for $x$ large enough, then $H$ is
in the limit point case at infinity. In consequence, in the
present case $H$ is essentially self-adjoint if and only if it is
in the limit point case at zero.

In particular, for positive $V(x)$ ($g\geq 0$), if $V(x)\geq 3/4\
x^{-2}$ for $x$ sufficiently close to zero then $H$ is in the
limit point case at the origin. On the contrary, if $V(x)\leq
(3/4-\eps)\ x^{-2}$, for some $\eps>0$, then $H$ is in the limit
circle case at zero.

This confirms our results concerning the self-adjointness of $H$
in the different regions of the parameter $g$.}. In this region,
$H$ admits different self-adjoint extensions.

\section{Self-adjoint extensions of $H$ \label{SAE-H} }

Since $n_+=1=n_-$ for $1/2\leq \alpha < 3/2$, there exists a
one-parameter family of self-adjoint extensions of $H$, which are
in a one-to-one relationship with the isometries from ${\mathcal
K}_+$ onto $\mathcal{K}_-$ \cite{Reed-Simon}.

The deficiency subspaces ${\mathcal K}_+$ and ${\mathcal K}_-$
are generated by $\phi_+\equiv \phi_{\lambda=i}$ and $\phi_-
\equiv \phi_{\lambda=-i}=\phi_{+}^*$, respectively. Then,  each
isometry ${\mathcal U}_{\gamma}:{\mathcal K}_+\rightarrow
{\mathcal K}_-$ can be identified with the parameter $\gamma\in
[0,\pi)$ defined by
\begin{equation}
    {\mathcal U}_{\gamma}
    \phi_+ = e^{-2i\gamma}\phi_-.
\end{equation}

The corresponding self-adjoint operator, $H_{\gamma}$, is defined
on a dense subspace \cite{Reed-Simon}
\begin{equation}
    {\mathcal D}(H_{\gamma})\subset
    {\mathcal D}(H^{\dagger})={\mathcal D}(\overline{H})\oplus
    {\mathcal K}_+\oplus {\mathcal K}_- ,
\end{equation}
where $\overline{H}$ is the closure of $H$. Functions $\phi\in
{\mathcal D}(H_{\gamma})$ can be written as
\begin{equation}\label{phi-suma}
    \phi=\phi_0+A \left( \phi_+ + e^{-2i\gamma}\phi_- \right),
\end{equation}
with $\phi_0\in {\mathcal D}(\overline{H})$ and $A$ a constant.
Since $H_{\gamma}$ is a restriction of $H^{\dagger}$, we have
\begin{equation}\label{H-gamma}
  H_{\gamma}\phi = H^{\dagger} \phi =
  \overline{H}\phi_0 + i A \left( \phi_+ - e^{-2i\gamma}\phi_-
  \right).
\end{equation}

\bigskip

In the following we take $g\geq 0\Rightarrow 1\leq \alpha <3/2$.
As we will see, condition (\ref{phi-suma}) determines the behavior
of $\phi\in {\mathcal D}(H_{\gamma})$ near the origin. Taking the
logarithmic derivative of $\phi$ we get,
\begin{equation}\label{vNc}
    \frac{\phi'}{\phi}=\frac{e^{i\gamma}\phi_0'+ 2A\,
    \Re\left(e^{i\gamma}\phi_+'\right)}
    {e^{i\gamma}\phi_0+2A\,
    \Re\left(e^{i\gamma}\phi_+\right)}.
\end{equation}
In this expression, the terms coming from $\phi_+$ give the
leading contributions for small $x$. In fact, in Appendix
\ref{closure} we show that  $\phi_0(x)=o(x^{\alpha})$ and
$\phi_0'(x)=o(x^{\alpha-1})$. Then, for the right hand side of
eq.\ (\ref{vNc}) we get \cite{Abramowitz} (see eq.\
(\ref{eigfun})),
\begin{equation}\label{bc}\begin{array}{c}\displaystyle{
   \frac{\phi'(x)}{\phi(x)}=\frac{1-\alpha}{x}+
   (2\alpha-1)\frac{\Gamma (\frac{1}{2}-\alpha)}
   {\Gamma (\alpha-\frac{1}{2})} \times
   }\\ \\
   \displaystyle{
   \frac{\cos{(\gamma-\gamma_1)}}
   {\cos{(\gamma-\gamma_2)}}\cdot x^{2\alpha-2}+
   o(x^{2\alpha-2})},
\end{array}
\end{equation}
where we have called
$\gamma_1=\arg\left\{\Gamma[(-2\alpha+3-i)/4]\right\}$ and
$\gamma_2=\arg\left\{\Gamma[(2\alpha+1-i)/4]\right\}$.

Thus, the limit of eq.\ (\ref{vNc}) for $x\to 0^+$ gives the
appropriate boundary condition for the functions in the domain of
the particular SAE . As we will see, this boundary condition will
finally determine a discrete spectrum for $H_{\gamma}$.

\section{The spectrum} \label{spectrum-SAE}

The boundary condition specified in eq.\ (\ref{bc}) characterizes
the domain of a particular SAE of the operator $H$, $H_\gamma$. In
order to determine its spectrum, we must find the solutions of
(\ref{eigequ}), $\phi_{\lambda}$ as given in (\ref{eigfun}) with
$\lambda\in\mathbf{R}$, which satisfy this boundary condition.
Their behavior near the origin is given by (see eq.\
(\ref{eigfun}) and \cite{Abramowitz}),
\begin{equation}\label{eigatori}\begin{array}{c}
  \displaystyle{\frac{\phi_{\lambda}'(x)}{\phi_{\lambda}(x)}=
  \frac{1-\alpha}{x}+(2\alpha-1)
  \frac{\Gamma (\frac{1}{2}-\alpha)}
  {\Gamma (\alpha-\frac{1}{2})} \times} \\ \\
  \displaystyle{ \frac{\Gamma \left[\frac{2\alpha+1-\lambda}{4}\right]}
  {\Gamma \left[\frac{-2\alpha+3-\lambda}{4}\right]}\cdot
   x^{2\alpha-2}+ o(x^{2\alpha-2}). }
\end{array}
\end{equation}
Comparison of eqs.\ (\ref{bc}) and (\ref{eigatori}) immediately
leads us to
\begin{equation}\label{spe}
    \frac{\Gamma \left(\kappa-\frac{\lambda}{4}\right)}
    {\Gamma\left(1-\kappa-\frac{\lambda}{4}\right)}=
    \beta(\gamma,\kappa),
\end{equation}
where we have defined the parameters
\begin{equation}\label{param}
  \begin{array}{c}
  \displaystyle{
    \kappa=\frac{2\alpha+1}{4}=
    \frac 1 4 \left( 2+\sqrt{1+4g} \right) \in [3/4,1) }\\ \\
    \displaystyle{
    \beta(\gamma,\kappa)=\cos{(\gamma-\gamma_1)}/\cos{(\gamma-\gamma_2)}}.
  \end{array}
\end{equation}

Eq.\ (\ref{spe}) determines a discrete spectrum for each SAE. In
Figure 1.\ we plot both sides of eq.\ (\ref{spe}) as a function of
$\lambda$, for $\kappa=4/5$ and $\beta=1$. The absciss\ae\ of the
intersections of this two functions give the corresponding
spectrum.

Notice that each SAE can equivalently be characterized by
$\beta\in \mathbf{R}\cup \{-\infty\}$. Then, we will also use the
notation $H_{(\beta)}$ to design this SAE.

\begin{figure}\label{figspe}
    \epsffile{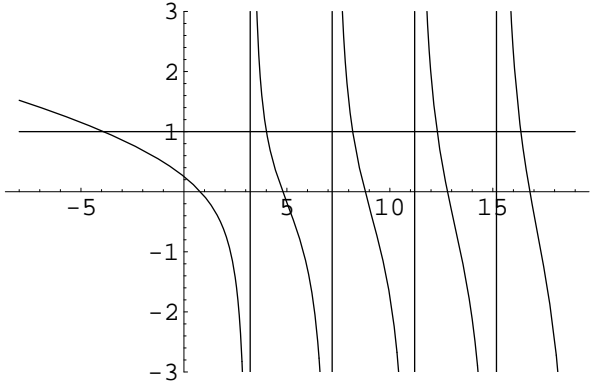}
    \caption{$F(\lambda)\equiv\frac{\Gamma
    \left(\kappa-\frac{\lambda}{4}\right)}
    {\Gamma\left(1-\kappa-\frac{\lambda}{4}\right)}$ as a function of
    $\lambda$, for $\kappa=4/5$. The solutions of $F(\lambda)=\beta$
    give the spectrum of the SAE  identified by $\beta$.}
\end{figure}

The spectrum of $H_{(\beta)}$ is bounded from below, and presents
a negative eigenvalue for those SAE characterized by
$\beta>\Gamma(\kappa)/\Gamma(1-\kappa)$ (even though the
potential $V(x)\geq 2 \sqrt{g}\geq 0$). Moreover, there is no
common lower bound; instead, any negative real is in the spectrum
of some SAE.

%%%%%%%%%%%%%%%%%%%%%%%%%%%%%%%%%

For any value of $g$, there are two particular SAE for which the
spectrum can be easily worked out (see eq.\ (\ref{spe})):
\begin{itemize}
\item {For $\beta=0$  the spectrum is given by
\begin{equation}\label{beta=0}
    \lambda_n=4(n+1-\kappa),
\end{equation}
with $n=0,1,2,\ldots$}
  \item {For $\beta= - \infty$  the spectrum
is given by
\begin{equation}\label{beta=-infty}
    \lambda_n=4(n+\kappa),
\end{equation}
with $n=0,1,2,\ldots$}
\end{itemize}

For other values of $\beta$, the eigenvalues grow linearly with
$n$,
\begin{equation}\label{linear-n}
  4(n-1+\kappa)< \lambda_{n}<4(n+\kappa).
\end{equation}

%%%%%%%%%%%%%%%%%%%%%%%%%%%%%%%%%

 {\small

 \subsection*{The case with $g=0$}

 It is instructive to consider the
particularly simple case of the harmonic oscillator in the
half-line, for which there is no singularity in the potential.
Indeed, for $g=0$ ($\alpha=1$ or $\kappa=3/4$), the boundary
condition (eq.\ (\ref{bc})) reads,
\begin{equation}
    \frac{\phi'(x)}{\phi(x)}=-2\beta + O(x)
\end{equation}
or, equivalently,
\begin{equation}
    \lim_{x\rightarrow
    0^+}\left\{\phi'(x)+2\beta\,\phi(x)\right\}=0,
\end{equation}
which corresponds to Robin boundary conditions at the origin.
Dirichlet and Neumann boundary conditions are obtained for
$\beta=-\infty$ and $\beta=0$, respectively.

Let's now study the eigenfunctions and eigenvalues of the
self-adjoint extensions of $H$ corresponding to different values
of $\beta$.

{\subsubsection*{ Dirichlet boundary conditions ($\beta=-\infty$)}
Since $\kappa=3/4$, the eigenvalues (see eq.\ (\ref{beta=-infty}))
are given by
\begin{equation}
    \lambda_n=4n+3,
\end{equation}
where $n=0,1,2,\ldots$

Since the Hamiltonian (eq.\ (\ref{ham})) corresponds in this case
to a particle with mass $m=1/2$ and frequency $\omega=2$, the
eigenvalues of this SAE can be written as
$\lambda_n=\omega[(2n+1)+1/2]$, which coincides with the spectrum
of the odd parity eigenvectors of the harmonic oscillator on the
complete real line.

In fact, the eigenfunctions are given by (see eq.\ (\ref{eigfun})
and \cite{Abramowitz}),
\begin{equation}
    \phi_n=2^{-2n-1}e^{-\frac{x^2}{2}}H_{2n+1}(x).
\end{equation}
}

{\subsubsection*{ Neumann boundary conditions ($\beta=0$)} In this
case, the eigenvalues (see eq.\ (\ref{beta=0})) are given by
\begin{equation}
    \lambda_n=4n+1,
\end{equation}
where $n=0,1,2,\ldots$. This eigenvalues can be written as
$\lambda_n=\omega(2n+1/2)$, which coincides with the even parity
sector of the harmonic oscillator spectrum on the complete real
line.

The eigenfunctions are now given by (see eq.\ (\ref{eigfun})),
\begin{equation}
    \phi_n=2^{-2n}e^{-\frac{x^2}{2}}H_{2n}(x).
\end{equation}
}

{\subsubsection*{ Robin boundary conditions ($\beta\neq
0,-\infty$)}

For finite $\beta \neq 0$, the eigenfunctions are given by (eq.\
(\ref{eigfun})),
\begin{equation}
    \phi_{\lambda}=xe^{-\frac{x^2}{2}}U\left(\frac{3-\lambda}{4};
    \frac{3}{2};x^2\right),
\end{equation}
and the corresponding eigenvalues are determined by the
trascendental equation
\begin{equation}
    \frac{\Gamma \left(\frac{3-\lambda}{4}\right)}
    {\Gamma\left(\frac{1-\lambda}{4}\right)}=
    \beta(\gamma,3/4).
\end{equation}
Notice that, for general Robin boundary condition, the ground
state is negative (less than the minimum of the potential) for
$\beta>\Gamma[\frac{3}{4}]/\Gamma[\frac{1}{4}]$. } }

%%%%%%%%%%%%%%%%%%%%%%%%%%%%%%%%%

\section{The integral representation for the $\zeta$-function }
\label{integ-rep}

The spectrum of each SAE of the operator $H$ in (\ref{ham}) is
determined by eq.\ (\ref{spe}), for any given
$\beta\in[-\infty,\infty)$. In this section, we will study the
pole structure of the associated $\zeta$-function, defined as
\begin{equation}\label{zeta-def}
  \zeta_\beta(s)\equiv Tr\left\{H_{(\beta)}^{-s}\right\} =
  \sum_{n} \lambda_{\beta,n}^{-s}.
\end{equation}
Notice that, since the eigenvalues grow linearly with $n$ (see
eq.\ (\ref{linear-n})), $\zeta_\beta(s)$ is analytic in the
half-plane $\Re(s)>1$.

For finite $\beta$, let us  define the holomorphic function,
\begin{equation}
    f(\lambda)=\frac{1}
    {\Gamma\left(1-\kappa-\frac{\lambda}{4}\right)}-\frac
    {\beta}{\Gamma \left(\kappa-\frac{\lambda}{4}\right)},
\end{equation}
with $\frac{3}{4}\leq \kappa<1$. The eigenvalues of the
self-adjoint operator $H_{(\beta)}$ correspond to the  zeroes of
$f(\lambda)$ which, consequently, are all real. They are also
positive, with the only possible exception of the first one,
according to the discussion in the previous Section.

Moreover, the zeroes of $f(\lambda)$ are simple. To prove this,
let's assume the converse is true, i.\ e.\ there is a $\lambda\in
\mathbf{R}$ such that $f(\lambda)=f'(\lambda)=0$. Taking into
account that
\begin{equation}\begin{array}{c}
  f'(\lambda)=\frac{\psi\left(1-\kappa-\frac{\lambda}{4}\right)}
        {4\Gamma\left(1-\kappa-\frac{\lambda}{4}\right)}-\beta
        \frac{\psi\left(\kappa-\frac{\lambda}{4}\right)}
        {4\Gamma\left(\kappa-\frac{\lambda}{4}\right)}= \\ \\
  =\frac{1}{4}
            \left\{\frac{\left[\psi\left(1-\kappa-\frac{\lambda}{4}\right)-
            \psi\left(\kappa-\frac{\lambda}{4}\right)\right]}
            {\Gamma\left(1-\kappa-\frac{\lambda}{4}\right)}+
            \psi\left(\kappa-\lambda/4\right)f(\lambda)
            \right\},
\end{array}
\end{equation}
where $\psi(z)$ is the polygamma function, we see that our
assumption requires that
\begin{equation}
    \psi\left(1-\kappa-\lambda/4\right)=
            \psi\left(\kappa-\lambda/4\right),
\end{equation}
which is not the case for any $\lambda\in \mathbf{R}$, if $\frac34
\leq \kappa <1$.

Therefore, the $\zeta$-function can be represented as the
integral on the complex plane
\begin{equation}\label{z-rep}
  \zeta_\beta(s)=\frac{1}{2\pi
  i}\oint_{C}\lambda^{-s}\frac{f'(\lambda)}{f(\lambda)}+
  \Theta(-\lambda_{0,\beta}) \lambda_{0,\beta}^{-s},
\end{equation}
where $C$ is a curve which encircles the positive zeroes of
$f(\lambda)$ counterclockwise. In eq.\ (\ref{z-rep}),
$\Theta(y)=1$ for $y>0$ and $\Theta(y)=0$ for $y\leq 0$.

Let us consider the dominant asymptotic behavior of the quotient
\begin{equation}\label{int}\begin{array}{c}\displaystyle{
  \frac{f'(\lambda)}{f(\lambda)} }= \\ \\ \displaystyle{ =
              \frac{\left[\psi\left(1-\kappa-\frac{\lambda}{4}\right)-
            \psi\left(\kappa-\frac{\lambda}{4}\right)\right]}
            {4\left(1-\beta\frac{\Gamma\left(1-\kappa-\frac{\lambda}{4}
            \right)}
            {\Gamma\left(\kappa-\frac{\lambda}{4}\right)}\right)}+
            \frac{1}{4}\,
            \psi\left(\kappa-\lambda/4\right)
             }.
\end{array}
\end{equation}
For $|\arg(-\lambda)|<\pi$ and $|\lambda|\rightarrow \infty$, it
is sufficient to write
\begin{eqnarray}
    \psi\left(\kappa-\lambda/4\right)=\log{(-\lambda)}+O(1),
    \label{fun1}\\ \nonumber \\
    \psi\left(1-\kappa-\lambda/4\right)-
    \psi\left(\kappa-\lambda/4\right)=O(\lambda^{-1}),
    \label{fun2}\\ \nonumber \\ \label{fun3}
    \frac{\Gamma\left(1-\kappa-\frac{\lambda}{4}\right)}
            {\Gamma\left(\kappa-\frac{\lambda}{4}\right)}=
            O(\lambda^{1-2\kappa}).
\end{eqnarray}

Consequently, for $\Re(s)>1$ the path of integration in
(\ref{z-rep}) can be deformed to a vertical line, to get
\begin{equation}\label{z-rep-imag}
  \zeta_\beta(s)=\frac{-1}{2\pi
  i}\int_{-i\infty+0}^{i\infty+0}\lambda^{-s}
  \frac{f'(\lambda)}{f(\lambda)}\, d\lambda+
  h(s),
\end{equation}
where $h(s)$ (the contribution of the negative eigenvalue, if any)
is a holomorphic function.

\section{Pole structure of the $\zeta$-function} \label{pole-structure}

The integral in eq.\ (\ref{z-rep-imag}) defines $\zeta_\beta(s)$
as an analytic function in the half-plane $\Re(s)>1$, which can be
meromorphically extended to the whole complex $s$-plane. It can be
written as
\begin{equation}\label{zetint}\begin{array}{c}
        \displaystyle{
        \zeta_\beta (s)=
        -\frac{1}{2\pi i}\int_{i}^{i\infty}
        \frac{f'(\lambda)}{f(\lambda)}\,
        \lambda^{-s}\,d\lambda\,- }\\ \\
        \displaystyle{
        -\frac{1}{2\pi i}
        \int_{-i\infty}^{-i} \frac{f'(\lambda)}{f(\lambda)}\,
        \lambda^{-s}\,d\lambda + h_1(s)=} \\ \\
        \displaystyle{
        =  -\frac{e^{-i s \pi/2}}{2\pi}  \int_{1}^{\infty}
        \frac{f'(i\mu)}{f(i\mu)}\,
        \mu^{-s}\,d\mu - }\\ \\
        \displaystyle{
        -\frac{e^{i s \pi/2}}{2\pi}
        \int_{1}^{\infty} \frac{f'(-i\mu)}{f(-i\mu)}\,
        \mu^{-s}\,d\mu + h_1(s)},
\end{array}
\end{equation}
where $h_1(s)$ is a holomorphic function.

\bigskip

In Appendix \ref{asymptotic} we work out the asymptotic expansion
of $f'(\lambda)/f(\lambda)$, which is given by
\begin{eqnarray}\label{te}\begin{array}{c}
    \displaystyle{
  \frac{f'(\lambda)}{f(\lambda)}\sim
    \frac{1}{4}\log{(-\lambda)}+
    \frac{1}{4}
    \sum_{k=0}^\infty c_k(\kappa)\,(-\lambda)^{-k}+ }\\
    \\
    \displaystyle{
    +\sum_{N=1}^\infty \sum_{n=0}^\infty
    C_{N,n}(\kappa,\beta)\,(-\lambda)^{-N(2\kappa-1)-2n-1}},
\end{array}
\end{eqnarray}
where the coefficients $c_k(\kappa)$ are polynomials in $\kappa$
whose explicit form is not needed for our purposes, and
\begin{equation}\label{CNn}\begin{array}{c}
  C_{N,n}(\kappa,\beta) =  \\ \\
  \displaystyle{
  =-\left( 4^{2\kappa-1} \beta \right)^N \left(
  {2\kappa-1+\frac{2n}{N} }\right) b_{n}(\kappa,N)},
\end{array}
\end{equation}
with $b_{n}(\kappa,N)$ defined in eq.\ (\ref{bs-as}) (see also
eq.\ (\ref{prim-b})).

As can be seen from eq.\ (\ref{te}), the asymptotic expansion of
$f'(\lambda)/f(\lambda)$ contains the logarithmic term $\frac 1 4
\log(-\lambda)$, and a series of non positive integer powers of
$\lambda$, both coming from the $\psi$-function in the last term
in the right hand side of (\ref{int}). There is also a series of
decreasing $\kappa$-dependent powers of $\lambda$, which comes
from the first term in the right hand side of (\ref{int}).

\bigskip

%%%%%%%%%%%%%%%%%%%%%%%%%%%%%%%%%%%

For the dominant logarithmic term we get from (\ref{zetint})
\begin{equation}\label{dominant}\begin{array}{c}
    \displaystyle{
  -\frac{1}{8 \pi }\int_{1}^{\infty}\left[e^{-i\frac{\pi s}{2}}
        \log{(e^{-i\frac{\pi}{2}}\mu)}+e^{i\frac{\pi s}{2}}
        \log{(e^{i\frac{\pi}{2}}\mu)}\
        \right]\mu^{-s}
        \,d\mu }\\ \\
        \displaystyle{
  = \frac{\sin (\frac{\pi \,s}{2})}{8\,\left(s -1  \right) } -
  \frac{\cos (\frac{\pi \,s}{2})}
  {4\,\pi \,{\left( s  -1 \right) }^2}
  =\frac{1}{4} \frac{1}{(s-1)}+ h_2(s)},
\end{array}
\end{equation}
where $h_2(s)$ is holomorphic. The analytic extension of this
term presents a unique simple pole at $s=1$, with a residue equal
to $1/4$.

The remaining terms in the asymptotic expansion of
$f'(\lambda)/f(\lambda)$ are of the form $A_j (-\lambda)^{-j}$,
for some $j\geq 0$  (see eq.\ (\ref{te})). Replacing this in eq.\
(\ref{zetint}) we get
\begin{equation}\label{next-terms}\begin{array}{c}
    \displaystyle{
  -\frac{A_j}{2\pi }\int_{1}^{\infty}\left[
        e^{-i\frac{\pi }{2}(s-j)}
        +e^{i\frac{\pi }{2}(s-j)}
        \right]\, \mu^{-s-j}
        \,d\mu= }\\ \\
        \displaystyle{
  =-\frac{A_j}{\pi }
  \cos\left(\frac{\pi}{2} (s-j)\right) \ \frac{1}{s-(1-j)}= }\\ \\
  \displaystyle{
  =-\frac{A_j\, \sin(\pi j)}{\pi}\,
  \frac{1}{s-(1-j)}+ h_3(s)},
\end{array}
\end{equation}
where $h_3(s)$ is holomorphic.

So, from each power dependent term in the asymptotic expansion of
$f'(\lambda)/f(\lambda)$, proportional to  $(-\lambda)^{-j}$, we
get a unique simple pole at $s=1-j$, with a residue given by
$\displaystyle{- (A_j/\pi)\, \sin(\pi j)}$.

Notice that this residue vanishes for integer values of $j$. In
particular, this is the case for all the contribution coming from
the asymptotic expansion of $\psi(\kappa - \lambda/4)$ in the last
term in the right hand side of eq.\ (\ref{int}), except for the
first one, the logarithmic term leading to eq.\ (\ref{dominant}).
In fact, this is the only singularity present in the
$\beta=-\infty$ and $\beta=0$ cases (see eq.\ (\ref{int})).

But in general, for $\frac{3}{4}\leq \kappa<1$, there are also
poles at non integer values of $s$, as follows from (\ref{te}).

\bigskip

In conclusion, besides the pole at $s=1$ with residue $1/4$,  for
each pair of integers
\begin{equation}\label{Nn}
  (N,n),\  {\rm with}\  N=1,2,3,\ldots,\ {\rm and}\
  n=0,1,2,\ldots,
\end{equation}
the $\zeta$-function of the SAE of $H$ characterized by the
parameter $\beta$, $\zeta_\beta(s)$, has a contribution with a
simple pole at the negative value
\begin{equation}\label{pop}
    s=-N(2\kappa-1)-2n\in(-N-2n,-\frac N 2 -2n],
\end{equation}
with a $\beta$-dependent residue given by
\begin{equation}\label{res}\begin{array}{c}
    \displaystyle{
  \left. {\rm Res}\,( \cdot )\right|_{s=-N(2\kappa-1)-2n} =}
  \\ \\     \displaystyle{ =
    \frac{(-1)^{N} }{\pi} \, C_{N,n}(\kappa,\beta)\, \sin(2\pi N\kappa)}.
\end{array}
\end{equation}

This is our main result, establishing the existence of
$\kappa$-dependent poles of the $\zeta$-function which, in
general,  are not located at a half an integer value of s.
Moreover, the residues depend on the SAE considered.

Finally, notice that when $\kappa$ is a rational number, there can
be several (but a finite number of) pairs $(N,n)$ contributing to
the same pole. They must satisfy
\begin{equation}\label{rational}
  \frac{n-n'}{N-N'}=\frac 1 2 -\kappa = - \frac p q \in
  \left(- 1/2,- 1/4\right],
\end{equation}
where $p,q\in \mathbf{N}$.

On the contrary, when $\kappa$ is irrational the poles coming from
different pairs $(N,n)$, also irrational, are not coincident.

\subsection{Poles and residues of $\zeta_\beta(s)$}

Let us recall that the logarithmic term in the expansion
(\ref{te}) leads to a pole at $s=1$ (see eq.\ (\ref{dominant}))
with a residue given by
\begin{equation}
   \left.{\rm Res}\,(\cdot)\right|_{s=1}=\frac{1}{4},
\end{equation}
independently of the SAE considered.

The other poles can be organized in sequences characterized by the
integer $N=1,2,\ldots$ In each sequence, successive poles differ
by $-2$.

For example, the poles corresponding to the pairs $(N=1,n)$, with
$n = 0,1,2,\ldots$, are located at (see eq.\ (\ref{pop}))
\begin{equation}
   -1 -2n<s=1-2\kappa -2n \leq -\frac 1 2 -2n,
\end{equation}
and have residues given by
\begin{equation}
  \left.{\rm Res}\,(\cdot)\right|_{s=1-2\kappa-2n}=-\frac{C_{1,n}
  (\kappa,\beta)}{\pi}
  \,  \sin(2\pi\kappa).
\end{equation}

Similarly, the poles arising from the $(N=2,n\geq 0)$ terms in the
asymptotic expansion  (\ref{te}) are located at
\begin{equation}
    -2-2n <s=2-4\kappa-2n<-1-2n,
\end{equation}
and have residues given by
\begin{equation}
    \left.{\rm Res}\,(\cdot)\right|_{s=2-4\kappa-2n}=
    \frac{C_{2,n}(\kappa,\beta)}{\pi}
    \,  \sin(4\pi\kappa).
\end{equation}

Notice that the poles in the $N$-th sequence have residues
proportional to $\beta^N$ (see eq.\ (\ref{CNn})).

\bigskip

Finally, let us stress that a pole of $\zeta_\beta(s)$ at a non
integer $s=-N(2\kappa-1)-2n$, as in (\ref{pop}), implies the
presence of a term in the small-$t$ asymptotic expansion of
Tr$\displaystyle{\left\{e^{-t\, H_{(\beta)}}\right\}}$ of the form
\begin{equation}\label{heat}
 {A_{[N(2\kappa-1)+2n]}}\,
  t^{N(2\kappa-1)+2n},
\end{equation}
with a coefficient related to the residue by
\begin{equation}\label{heat-coef}\begin{array}{c}
  A_{[N(2\kappa-1)+2n]}= \\ \\
  =\Gamma(-N(2\kappa-1)-2n)\,
    \left. {\rm Res}\, \zeta_\beta(s)
    \right|_{s=-N(2\kappa-1)-2n}.
\end{array}
\end{equation}

\subsection{$\zeta$-function singularities from the asymptotic
expansion of the eigenvalues }

 The singular behavior found for $\zeta_\beta(s)$ can be
confirmed (at least for the first few poles) by determining from
(\ref{spe}) the asymptotic expansion of the eigenvalues
$\lambda_{\beta,n}$ for $n\gg 1$. Indeed, one can make the Ansatz
\begin{equation}\label{ansatz}
    \frac{\lambda_{\beta,n}}{4}=1-\kappa+n+\varepsilon,
\end{equation}
and self-consistently determine $\varepsilon$ through successive
corrections. For the first terms we get
\begin{equation}\label{asymp-eigen}\begin{array}{c}
    \displaystyle{
   \frac{\lambda_{\beta,n}}{4}\sim 1-\kappa+n +
   \frac{\beta }
  {\pi }\,\sin (2\,\pi \,\kappa )\, n^{1 - 2\,\kappa } +
   } \\ \\
  \displaystyle{ +
  \frac{\beta }{
    \pi } \,\left( 1 - 3\,\kappa  + 2\,{\kappa }^2
      \right) \,\sin (2\,\pi \,\kappa )\, n^{-2\,\kappa } - }\\ \\
  \displaystyle{-\frac{ {\beta }^2 }{2\,\pi }\,
      \sin (4\,\pi \,\kappa )  \, n^{2 - 4\,\kappa }+\dots},
\end{array}
\end{equation}
where we have retained only powers of $n$ greater than $-2$. This
leads, for the $\zeta$-function in eq.\ (\ref{zeta-def}), to
\begin{equation}\begin{array}{c}
    \displaystyle{
  \zeta_\beta(s)
  \sim 4^{-s}\, \zeta(s)+s\, 4^{-s}\, (\kappa-1)\, \zeta( s+1)\, +} \\ \\
  \displaystyle{+  s\,
    \left(s + 1 \right)\, 4^{-s}
  \frac{{\left(\kappa -1  \right) }^2\, }{2}\,\zeta( s+2 ) - }\\ \\
  \displaystyle{
  -s\,4^{-s}\, \frac{\beta }{\pi }\,
      \sin \left(  2\,\kappa  \,\pi\right) \,
      \zeta(s+2\,\kappa ) - }\\ \\
      \displaystyle{ -s\,
      \left( s + 2\,\kappa  \right)4^{-s}\, \frac{\beta }{
      \pi } \,\left( \kappa  -1  \right)  \,
      \sin (2\,\pi \,\kappa )\,
      \zeta(1 + s + 2\,\kappa )\, + }\\ \\
      \displaystyle{ + s\, 4^{-s}\,\frac{{\beta }^2}{2\,\pi } \,
    \sin (4\,\pi \,\kappa )\,
    \zeta( s-1  + 4\,\kappa ) \, +   \dots,}
\end{array}
\end{equation}
where $\zeta(z)$ is the Riemann $\zeta$-function, which presents a
unique simple pole at $z=1$, with a residue equal to $1$.

This result shows a pole structure in agreement with the one
previously described.

\section{Particular cases} \label{particular}

In this Section we will show how our results reduce to the usual
ones for $g=0$ (when there is no singularity in the potential). We
will also show that, for $\beta=0$ and $\beta=-\infty$, the
$\zeta$-function presents a unique simple pole.

\subsection{The $\beta=0$ and $\beta=-\infty$ SAE}

The $\zeta$-function for the SAE characterized by $\beta=0$  and
$\beta=-\infty$ can be exactly evaluated, since in these cases the
spectrum was explicitly computed in eqs.\ (\ref{beta=0}) and
(\ref{beta=-infty}) respectively. We get
\begin{equation}\label{zetas}
  \begin{array}{c}
  \displaystyle{
    \zeta_0(s)=4^{-s} \sum_{n=0}^\infty
    (n+1-\kappa)^{-s}=4^{-s}\zeta(s,1-\kappa),}\\ \\
  \displaystyle{
    \zeta_{-\infty}(s)=4^{-s} \sum_{n=0}^\infty
    (n+\kappa)^{-s}=4^{-s}\zeta(s,\kappa),}
  \end{array}
\end{equation}
where $\zeta(s,q)$ is the Hurwitz $\zeta$-function, whose
analytic extension presents only a simple pole at $s=1$, with a
residue Res\,$\zeta(s,q)|_{s=1}=1$. This leads, in both cases, to
a unique simple pole for the $\zeta$-function at $s=1$, with a
residue equal to $1/4$, in agreement with eq.\ (\ref{dominant}).

In fact, from eqs.\ (\ref{res}) and (\ref{CNn}) it is evident that
all the residues corresponding to negative poles vanish for
$\beta=0$. On the other hand, for $\beta=-\infty$,
$f'(\lambda)/f(\lambda)$ reduces to $\frac 1 4
\psi(\kappa-\lambda/4)$ (see eq.\ (\ref{int})), and the only term
leading to a singularity is the logarithm in the asymptotic
expansion (\ref{te}), as already discussed (see eq.\
(\ref{dominant})).

\subsection{The harmonic oscillator in the half-line}

For the harmonic oscillator in the half-line ($g=0$ or
$\kappa=3/4$) we still find a simple pole at $s=1$, with residue
$1/4$ (the only singularity for Dirchlet or Neumann boundary
conditions, as previously discussed).

For finite $\beta$, the remaining singularities are located at
(see eq.\ (\ref{pop})),
\begin{equation}
    s=-\frac{N}{2}-2n,\ N=1,2,3,\ldots,\ n=0,1,2,\ldots,
\end{equation}
with residues given by (see eq.\ (\ref{res})),
\begin{equation}\begin{array}{c}
    \displaystyle{
  \left. {\rm Res}\,( \cdot )\right|_{s=-\frac{N}{2}-2n} = }\\ \\
  \displaystyle{
  =\frac{(-1)^{N}}{\pi}\, \,C_{N,n}\left(\kappa =
    3/4,\beta \right) \, \sin\left(\frac{3\,\pi }{2}N\right) },
\end{array}
\end{equation}
which vanish for even $N$.

Then, each pole (except for the first one, at $s=1$) corresponds
to a negative half-integer,
\begin{equation}
  s=-k-1/2,\ k=0,1,2.\ldots
\end{equation}
Moreover, it is clear that for finitely many pairs $(N,n)$
satisfying $N+4n=2k+1$, the corresponding poles lie at the same
point.

Therefore, the residue of $\zeta_\beta(s)$ at $s=-k-1/2$ must be
computed by adding all these contributions, characterized by
$N=2(k-2n)+1$, with $n=0,1,2,\dots,[k/2]$. We get
\begin{equation}\label{suma-residuos}\begin{array}{c}
  \displaystyle{
  \left. {\rm Res}\,( \zeta_\beta(s) )\right|_{s= -k - \frac 1 2} =} \\ \\
  \displaystyle{ =
   \frac{(-1)^{k+1
   }}{\pi}\,\sum_{n=0}^{[k/2]}  \,C_{[2(k-2n)+1],n}\left(
    \kappa=3/4,\beta \right)  }.
\end{array}
\end{equation}

For example, for $k=0$, the residue is
\begin{equation}
  \left. {\rm Res}\,( \zeta_\beta(s) )\right|_{s= - \frac 1 2} =
  - \frac{1}{\pi} \,C_{1,0}\left(
    \kappa=3/4,\beta \right)=\frac{\beta }{\pi}
\end{equation}
and, for $k=1$,
\begin{equation}
  \left. {\rm Res}\,( \zeta_\beta(s) )\right|_{s= - \frac 3 2} =
  \frac{1}{\pi} \,C_{3,0}\left(
    \kappa=3/4,\beta \right)=-\frac{4}{\pi}\,\beta^3.
\end{equation}

\section{Conclusions} \label{conclusions}

In this article we have analyzed the pole structure of the
$\zeta$-function of the Hamiltonian describing a quantum Schr\oe
dinger particle living in the half-line $\mathbf{R}^+$, subject to
the singular potential $V(x)=g x^{-2}+ x^2$.

We have specified the domain of the adjoint of the Hamiltonian,
$H^\dagger$, and determined the deficiency subspaces of $H$,
initially defined on ${\mathcal C}_0^\infty(\mathbf{R^+})$. We
have shown that, for $-1/4\leq g<3/4$, $H$ admits nontrivial
self-adjoint extensions which depend on a continuous real
parameter $\beta$.

For computational convenience, we have limited our analysis to the
range $0\leq g<3/4$.

Once determined the closure of $H$ (studied in Appendix
\ref{closure}), we were able to characterize each SAE
$H_{(\beta)}$ by the behavior (singular, in general) near the
origin of the functions in the corresponding domain of definition.
This relation also allowed for the identification of the spectrum
of $H_{(\beta)}$ with the zeroes of an analytic function
$f(\lambda)$.

The asymptotic expansion of $f(\lambda)$ (outlined in Appendix
\ref{asymptotic}) led to the determination of the poles and
residues of the $\zeta$-function associated with $H_{(\beta)}$,
$\zeta_\beta(s)$.

We have shown that the poles of $\zeta_\beta(s)$ can be organized
in sequences characterized by an integer $N=1,2,3,\dots$, and are
located at $s=-N(2\kappa-1)-2n$, with $n=0,1,2,\dots$ and
$\kappa=(1+\sqrt{g+1/4})/2\in[3/4,1)$. Notice that these values of
$s$ are not, in general, half an integer (which are the expected
positions of the poles for a second order differential operator
with smooth coefficients on a compact segment), and they are
irrational numbers for irrational values of $\kappa$.

We have also found that the residues depend on the parameter
$\beta$ characterizing the SAE $H_{(\beta)}$.

We have confirmed this $\zeta$-function pole structure (for the
first poles) through the comparison with the results obtained from
the asymptotic behavior of the eigenvalues.

These results also imply  that the small-$t$ asymptotic expansion
of the heat kernel of $H_{(\beta)}$ contains  powers of $t$ which
(in general) are not half an integer, and that the corresponding
coefficients depend on the SAE.

Finally, several particular cases were analyzed, finding that our
results are consistent with the known ones. In particular, for the
harmonic oscillator ($g=0$) in the half line, subject to any local
boundary condition at $x=0$, the poles lie at half-integer values
of $s$.

\bigskip

A final remark is in order: Notice that the unusual pole structure
previously described is a consequence of having a potential with a
{\it moderate} singular behavior near the origin. In fact, for $g
\geq 3/4$, where the Hamiltonian $H$ is essentially self-adjoint
due to a stronger singular behavior of $V(x)$, the
$\zeta$-function simply reduces  to $4^{-s} \, \zeta(s,\kappa)$
(see eq.\ (\ref{spec-ESA})), which presents a unique pole at $s=1$
with residue $1/4$.

\bigskip

A similar pole structure is obtained for the Hamiltonian
$\zeta$-function of charged Dirac particles living in
$(2+1)$-dimensions, in the presence of both a uniform magnetic
field and a singular magnetic tube with a non-integer flux. This
problem was considered in \cite{H-P}, where it was shown that the
Hamiltonian restricted to a {\it critical} angular momentum
subspace admits nontrivial SAE, whose spectra are determined by a
trascendental equation similar to (\ref{spe}). These results will
be reported elsewhere.

\section*{Acknowledgements}

We thank E.M.\ Santangelo and M.A.\ Muschietti for useful
discussions.

The authors acknowledge support from Fundaci\'{o}n Antorchas and DAAD
(grant 13887/1-87).

HF and PAGP also acknowledge support from CO\-NI\-CET (grant
0459/98) and CIC-PBA (Argentina).

\appendix

\section{Closure of $H$} \label{closure}

In this Section we will justify to disregard the contributions
from the functions in the domain of the closure $\overline{H}$ to
the boundary condition, eq.\ (\ref{bc}). Indeed, we will show that
if $\phi\in{\mathcal D}(\overline{H})$ then
\begin{equation}
    \phi(x)=o(x^{\alpha}) \quad {\rm and}\quad
    \phi'(x)=o(x^{\alpha-1})
\end{equation}
near the origin, for any $\alpha<3/2$.

In order to determine the closure of the Hamiltonian's graph we
must consider those Cauchy sequences in ${\mathcal
D}(H)={\mathcal C}_0^\infty (\mathbf{R^+})$, $\{\varphi_n\}_{n\in
\mathbf{N}}$, such that $\{H\varphi_n\}_{n\in \mathbf{N}}$ are
also Cauchy sequences. Notice that, since the coefficients in $H$
are real (see eq.\ (\ref{ham})), we can consider real functions.

Let us call $\varphi=\varphi_n - \varphi_m$, with $n,m\in
\mathbf{N}$. Then $\varphi \rightarrow 0$ and $H\varphi
\rightarrow 0$ as $n,m \rightarrow \infty$.

Consider first the scalar product
\begin{equation}\label{phi-Hphi}\begin{array}{c}
  \displaystyle{\left( \varphi,H\varphi \right) =
  \int_0^\infty \varphi \left( - \varphi'' + \frac{g}{x^2}\, \varphi +
  x^2 \varphi \right)\, dx = } \\ \\
  \displaystyle{=  \int_0^\infty \left( \varphi'^2 + \frac{g}{x^2}\,
  \varphi^2 +
  x^2 \varphi^2 \right)\, dx  \leq ||\varphi|| \, ||H\varphi||
  \rightarrow 0}
\end{array}
\end{equation}
for $n,m \rightarrow \infty$. Therefore, for $g > 0$, we conclude
that
\begin{equation}\label{C-seq}
  \{\varphi'_n(x)\}_{n\in \mathbf{N}},\
  \displaystyle{\left\{\frac{\varphi_n(x)}{x}\right\}_{n\in
    \mathbf{N}}} \  {\rm and} \ \{x\, \varphi_n(x)\}_{n\in \mathbf{N}}
\end{equation}
are also Cauchy sequences.

\bigskip

We will now prove the following

\noindent{\bf Lemma:} Let $\{\varphi_n\}_{n\in \mathbf{N}}$ be a
Cauchy sequence in ${\mathcal D}(H)={\mathcal
C}_0^\infty(\mathbf{R^+})$ such that, for $g>0$, $1\leq a<2$ and
$g\neq(a^2 -1)/4$,
\begin{equation}
  \{H\varphi_n\}_{n\in \mathbf{N}},\
  \displaystyle{\left\{\frac{\varphi_n(x)}{x^a}\right\}_{n\in
  \mathbf{N}}}, \ {\rm and} \
  \displaystyle{\left\{\frac{\varphi_n'(x)}{x^{a-1}}\right\}_{n\in
  \mathbf{N}}}
\end{equation}
are also Cauchy sequences. Then,
\begin{equation}
  \displaystyle{\left\{\frac{\varphi_n(x)}{x^{1+{a}/{2}}}\right\}_{n\in
    \mathbf{N}}} \ {\rm and} \
    \displaystyle{\left\{\frac{\varphi_n'(x)}{x^{{a}/{2}}}\right\}_{n\in
    \mathbf{N}}}
\end{equation}
are Cauchy sequences too.

\noindent {\bf Proof:} As before, let $\varphi=\varphi_n -
\varphi_m$. First notice that, for $1\leq a<2$,
\begin{equation}\label{l0}\begin{array}{c}
  \displaystyle{\int_0^\infty \left( x^{1-a/2}\, \varphi(x) \right)^2\,
  dx \leq
  \int_0^1 \left(  \varphi(x) \right)^2\, dx + } \\ \\
  \displaystyle{+  \int_1^\infty  \left( x \, \varphi(x) \right)^2\,
  dx} \leq || \varphi(x) ||^2 + || x\,\varphi(x) ||^2.
\end{array}
\end{equation}
Then, from  (\ref{C-seq}), we see that $\displaystyle{ \left\{
x^{1-a/2}\, \varphi_n(x) \right\}_{n\in \mathbf{N}} }$ is also a
Cauchy sequence.

A straightforward calculation shows that
\begin{equation}\label{l1}\begin{array}{c}
  \displaystyle{ \left(\frac{\varphi(x)}{x^a}, H \varphi(x) \right) =
  \int_0^\infty
  \left\{ \left(\frac{\varphi'(x)}{x^{a/2}}\right)^2 + \right.} \\ \\
  \displaystyle{\left. + \left[ g - \frac{a(a+1)}{2} \right]
  \left(\frac{\varphi(x)}{x^{1+a/2}}\right)^2 +
  \left( x^{1-a/2}\, \varphi(x) \right)^2\right\}\, dx}.
\end{array}
\end{equation}
Similarly,
\begin{equation}\label{l2}\begin{array}{c}
  \displaystyle{ \left(\frac{\varphi'(x)}{x^{a-1}},
  H \varphi(x) \right) = \int_0^\infty
  \left\{-\left(\frac{a-1}{2} \right)
  \left(\frac{\varphi'(x)}{x^{a/2}}\right)^2 + \right.} \\ \\
  \displaystyle{\left.  g \left( \frac{a+1}{2} \right)
  \left(\frac{\varphi(x)}{x^{1+a/2}}\right)^2 + \left( \frac{a-3}{2} \right)
  \left( x^{1-a/2}\, \varphi(x) \right)^2\right\}\, dx}.
\end{array}
\end{equation}

Now, taking into account that the sum of fundamental sequences is
also a Cauchy sequence, we see that
\begin{equation}\label{l3}
  \left(A\,\frac{\varphi(x)}{x^a}+ B\, \frac{\varphi'(x)}{x^{a-1}}\, ,
  H \varphi(x) \right) \rightarrow 0
\end{equation}
when $n,m\rightarrow \infty$, for any pair of real numbers $A$
and $B$.

The coefficients of
$\displaystyle{\left(\frac{\varphi'(x)}{x^{a/2}}\right)^2}$ and
$\displaystyle{\left(\frac{\varphi(x)}{x^{1+a/2}}\right)^2}$ in the
expression of the scalar product in (\ref{l3}) are
\begin{equation}\label{l4}
  \begin{array}{c}
    \displaystyle{A -B\left(\frac{a-1}{2} \right)} \quad {\rm and} \\ \\
     \displaystyle{
     A \left(g- \frac{a(a+1)}{2} \right)+B\, g \left( \frac{a+1}{2}
     \right)}
  \end{array}
\end{equation}
respectively. One can see that, by choosing one of them as zero,
the other is non vanishing (except for $g=(a^2 -1)/4$).

It is easily seen that these results prove the Lemma.

\bigskip

For $a=1$, from (\ref{C-seq}) and the Lemma, we conclude that
\begin{equation}\label{l5}
  \displaystyle{\left\{\frac{\varphi_n(x)}{x^{3/{2}}}\right\}_{n\in
    \mathbf{N}}} \quad {\rm and} \quad
    \displaystyle{\left\{\frac{\varphi_n'(x)}{x^{{1}/{2}}}\right\}_{n\in
    \mathbf{N}}}
\end{equation}
are Cauchy sequences.

Let us first suppose that $g$ is an irrational number. Then,
applying iteratively the Lemma from (\ref{l5}) one can show that,
for any positive integer $k$,
\begin{equation}\label{l6}
  \displaystyle{\left\{\frac{\varphi_n(x)}{x^{2\left[ 1- \left(
  1/2 \right)^k \right]}}\right\}_{n\in
    \mathbf{N}}} \quad {\rm and} \quad
    \displaystyle{\left\{\frac{\varphi_n'(x)}{x^{2\left[ 1- \left(
  1/2 \right)^k \right]-1}}\right\}_{n\in
    \mathbf{N}}}
\end{equation}
are Cauchy sequences.

Finally, for any given $\varepsilon >0$ there are integers $k_1$
and $k_2$ such that $(1/2)^{k_1} \leq \varepsilon \leq
(1/2)^{k_2}$. Taking into account that
\begin{equation}\label{l7}
  \begin{array}{c}
    \displaystyle{  \frac{1}{x^{2-\varepsilon}} \leq
    \frac{1}{x^{2\left[ 1- \left(
  1/2 \right)^{k_1} \right]}} , \quad {\rm for} \ 0<x\leq 1},\\ \\
    \displaystyle{\frac{1}{x^{2-\varepsilon}} \leq
    \frac{1}{x^{2\left[ 1- \left(
  1/2 \right)^{k_2} \right]}} ,\quad {\rm for}\ x\geq 1},
  \end{array}
\end{equation}
one immediately  concludes that
$\displaystyle{\left\{\frac{\varphi_n(x)}{x^{2-\varepsilon}}\right\}_{n\in
\mathbf{N}}}$ is a Cauchy sequence.

A similar conclusion is easily obtained for
$\displaystyle{\left\{\frac{\varphi_n'(x)}{x^{1-\varepsilon}}\right\}_{n\in
\mathbf{N}}}$.

\bigskip

Let us now suppose that $g$ is a rational number. Then, from
(\ref{C-seq}) and (\ref{l5}) it is seen that we can choose an
irrational $a\in(1,3/2)$ from which the Lemma can also be applied
iteratively to arrive to the same conclusions.

\bigskip

In the following we will consider the behavior of the functions
near the origin.

For any $\varepsilon>0$, we can write
\begin{equation}\label{1}\begin{array}{c}
  \displaystyle{ x^{-\alpha} \, \varphi(x) =
  \int_0^x \left( y^{-\alpha}  \, \varphi(y)
  \right)'\, dy = } \\ \\
  \displaystyle{=\int_0^x y^{{-\alpha} +1-\varepsilon}\left\{ {-\alpha}
  \,   \frac{\varphi(y)}{y^{2-\varepsilon}} +
  \frac{\varphi'(y)}{y^{1-\varepsilon}} \right\} \, dy .}
\end{array}
\end{equation}
So, for $x\leq 1$, ${\alpha} < 3/2$ and $\varepsilon$ small
enough, we have
\begin{equation}\label{2}\begin{array}{c}
  \displaystyle{\left| x^{-\alpha}  \, \varphi(x) \right|\leq
  \left(
  \int_0^1 y^{2({-\alpha} +1-\varepsilon)}dy \right)^{1/2}
  \left\{ |{\alpha} |\left|\left|
  \frac{\varphi(y)}{y^{2-\varepsilon}} \right|\right| \right.+} \\ \\
  \displaystyle{ \left. +
  \left|\left|
  \frac{\varphi'(y)}{y^{1-\varepsilon}} \right|\right|  \right\}
  \rightarrow_{n,m \rightarrow \infty} 0.}
\end{array}
\end{equation}
Therefore, the sequence $\{x^{-\alpha}  \,
\varphi_n(x)\}_{n\in\mathbf{N}}$, with ${\alpha}<3/2$, is uniformly
convergent in $[0,1]$, and its limit is a continuous function
vanishing at the origin,
\begin{equation}\label{3}
  \displaystyle{\lim_{n\rightarrow\infty}\left( x^{-\alpha}
  \,\varphi_n(x) \right)=
  x^{-\alpha}  \,\phi(x),}
\end{equation}
\begin{equation}\label{4}
  \displaystyle{\lim_{x\rightarrow 0^+} \left(x^{-\alpha}
  \,\phi(x)\right)=0.}
\end{equation}
In particular, for ${\alpha}  = 0$ we have the uniform limit
\begin{equation}\label{4-5}
 \lim_{n\rightarrow\infty}\varphi_n(x) = \phi(x),
\end{equation}
which coincides with the limit of this sequence in
$\mathbf{L_2}(\mathbf{R^+})$.

\bigskip

On the other hand,  we can also write
\begin{equation}\label{6}\begin{array}{c}
  \displaystyle{\int_0^x y^{{-\alpha} +1}\,H\varphi(y)\,dy =
  -x^{{-\alpha} +1}\,\varphi'(x) + } \\ \\
  \displaystyle{+ \int_0^x y^{{-\alpha} +1-\varepsilon} \left\{
  ({-\alpha} +1)\,
  \frac{\varphi'(y)}{y^{1-\varepsilon}}+ g\,
  \frac{\varphi(y)}{y^{2-\varepsilon}} \right\}\, dy + } \\ \\
  \displaystyle{+ \int_0^x y^{{-\alpha} +2}\, y\, \varphi(y)\, dy}.
\end{array}
\end{equation}
Therefore, for $ x \leq 1$,  ${\alpha} < 3/2$ and $\varepsilon$
sufficiently small, we have
\begin{equation}\label{7}\begin{array}{c}
  \displaystyle{\left| x^{{-\alpha} +1}\,\varphi'(x) \right| \leq
  \left( \int_0^1 y^{2({-\alpha} +1)}\, dy \right)^{1/2} \left|\left|
  H\varphi(y) \right|\right| + } \\ \\
  \displaystyle{\left( \int_0^1 y^{2({-\alpha} +1-\varepsilon)}\,
  dy \right)^{1/2}
  \left\{ |{\alpha} -1| \left|\left| \frac{\varphi'(y)}
  {y^{1-\varepsilon}}
  \right|\right|+
  g \left|\left| \frac{\varphi(y)}{x^{2-\varepsilon}} \right|\right|
  \right\}+ }\\ \\
  \displaystyle{+ \left( \int_0^1 y^{2({-\alpha} +2)}\, dy \right)^{1/2}
  \left|\left|
  y\,\varphi(y) \right|\right| \rightarrow_{n,m\rightarrow \infty} 0.}
\end{array}
\end{equation}
Consequently, the sequence
$\{x^{{-\alpha}+1}\,\varphi_n'(x)\}_{n\in\mathbf{N}}$, with
${\alpha}< 3/2$, is uniformly convergent in $[0,1]$, and its limit
is a continuous function vanishing at the origin, which we write
as $x^{{-\alpha} +1} \,\chi(x)$:
\begin{equation}\label{8}
  \displaystyle{\lim_{n\rightarrow\infty}\left(
  x^{{-\alpha} +1}\,\varphi_n'(x) \right)=
   x^{{-\alpha} +1} \,\chi(x),}
\end{equation}
\begin{equation}\label{9}
  \displaystyle{\lim_{x\rightarrow 0^+} \left( x^{{-\alpha} +1}
  \,\chi(x)\right)=0.}
\end{equation}
In particular, for ${\alpha} =1$ we have the uniform limit
\begin{equation}\label{10}
  \lim_{n\rightarrow\infty}\varphi'_n(x) = \chi(x),
\end{equation}
which coincides with the limit of this sequence in
$\mathbf{L_2}(\mathbf{R^+})$ (see (\ref{C-seq})).

Let us now show that $\chi(x)=\phi'(x)$. Indeed, for $x\leq 1$,
we have
\begin{equation}\label{11}\begin{array}{c}
  \displaystyle{ \left| \phi(x)-\int_0^x \chi(y)\,dy \right|\leq }
  \\ \\
  \displaystyle{\leq \left| \phi(x)-\varphi_n(x) \right| +
  \left| \int_0^x \left(\chi(y)-\varphi'_n(y)\right)\,dy
  \right|\leq } \\ \\
  \displaystyle{\leq \left| \phi(x)-\varphi_n(x) \right| + \left|\left|
   \chi-\varphi'_n \right|\right| \rightarrow_{n \rightarrow \infty}
   0}.
\end{array}
\end{equation}
So, $\phi(x)$ is a differentiable function whose first derivative
is $\chi(x)$.

\bigskip

Equations (\ref{4}) and (\ref{9}) imply that, given
$\varepsilon_1>0$  and $\alpha<3/2$,
\begin{equation}\label{5}
  \left| \phi(x) \right| < \varepsilon_1 \, x^{{\alpha} }\quad {\rm
  and}\quad \left| \phi'(x) \right| < \varepsilon_1 \, x^{{\alpha-1} }
\end{equation}
if $x<\delta$, for some  $\delta>0$ small enough. This proves our
assertion.

\section{Asymptotic expansions}\label{asymptotic}

In this appendix we will compute the asymptotic expansion for
$f'(\lambda)/f(\lambda)$ as given in eq.\ (\ref{int}).

The asymptotic expansion for the polygamma function appearing in
the right hand side of eq.\ (\ref{int}) can be easily obtained
from Stirling's formula \cite{Abramowitz},
\begin{equation}\label{asymp-psi}
      \psi\left(\kappa-\lambda/4\right)\sim\log{(-\lambda)}+
    \sum_{i=0}^\infty c_i(\kappa) (-\lambda)^{-k},
    \nonumber
\end{equation}
where the coefficients $c_i(\kappa)$ are polynomials in $\kappa$
which we will not need to explicitly know for our purposes.

On the other hand, taking into account (\ref{fun3}), we can write
asymptotically for the first term in the right hand side of eq.\
(\ref{int})
\begin{equation}\label{asymp}
    \begin{array}{c}
    \displaystyle{
    \frac{\left[\psi\left(1-\kappa-\frac{\lambda}{4}\right)-
            \psi\left(\kappa-\frac{\lambda}{4}\right)\right]}
            {1-\beta\frac{\Gamma\left(1-\kappa-\frac{\lambda}{4}\right)}
            {\Gamma\left(\kappa-\frac{\lambda}{4}\right)}}\sim }\\
            \\
   \sum_{N=0}^{\infty}\beta^N
            \left[\frac{\Gamma\left(1-\kappa-\frac{\lambda}{4}\right)}
            {\Gamma\left(\kappa-\frac{\lambda}{4}\right)}\right]^N
            \left[\psi\left(1-\kappa-\frac{\lambda}{4}\right)-
            \psi\left(\kappa-\frac{\lambda}{4}\right)\right]= \\
            \\
    =\sum_{N=0}^{\infty}{\beta^N}
    \left[\frac{\Gamma\left(1-\kappa-\frac{\lambda}{4}\right)}
            {\Gamma\left(\kappa-\frac{\lambda}{4}\right)}\right]^N \,
            4\,\frac{d}{d(-\lambda)}\log
            \left[\frac{\Gamma\left(1-\kappa-\frac{\lambda}{4}\right)}
            {\Gamma\left(\kappa-\frac{\lambda}{4}\right)}\right]=\\
            \\
    =\sum_{N=0}^{\infty}\frac{\beta^N}{N}\,4\,
            \frac{d}{d(-\lambda)}
            \left[\frac{\Gamma\left(1-\kappa-\frac{\lambda}{4}\right)}
            {\Gamma\left(\kappa-\frac{\lambda}{4}\right)}\right]^N.
    \end{array}
\end{equation}

From the Stirling's formula \cite{Abramowitz} we get
\begin{equation}\label{gqg}\begin{array}{c}
    \displaystyle{
       \log\left[\frac{\Gamma\left(1-\kappa-\frac{\lambda}{4}\right)}
            {\Gamma\left(\kappa-\frac{\lambda}{4}\right)}\right]\sim
    (1-2\kappa)\log(-\frac{\lambda}{4})\,+}\\ \\
        \displaystyle{
    +{\left\{\displaystyle{\sum_{m=1}^\infty a_m(\kappa)
  (-\lambda)^{-2m}
  }\right\}}},
\end{array}
\end{equation}
where the coefficients in the series are given by
\begin{equation}\label{am(kappa)}\begin{array}{c}
    \displaystyle{
  a_m(\kappa)=\frac{2^{4m-1}}{2m+1}\left\{
  \left[ (1-\kappa)^{2m} - \kappa^{2m} \right]+
  \left(\frac{\kappa-1/2}{m}\right) \times\right. }\\ \\
      \displaystyle{
  \times\left[ (1-\kappa)^{2m} + \kappa^{2m}  \right]+
  (2m+1) \sum_{p=1}^m \frac{B_{2p}}{p(2p-1)} \times }
  \\ \\    \displaystyle{
  \left. \times
  \left(\begin{array}{c}
    2m-1 \\
    2p-2
  \end{array}\right) \left[ \kappa^{2(m-p)+1} -
  (1-\kappa)^{2(m-p)+1} \right] \right\} }.
\end{array}
\end{equation}
Then,
\begin{equation}\label{GsGN}\begin{array}{c}
    \displaystyle{
  \left[\frac{\Gamma\left(1-\kappa-\frac{\lambda}{4}\right)}
            {\Gamma\left(\kappa-\frac{\lambda}{4}\right)}\right]^N\sim }
            \\ \\    \displaystyle{
  \sim \left(-\frac{\lambda}{4}\right)^{-N(2\kappa-1)}
            \sum_{n=0}^{\infty}b_n(\kappa,N)\,(-\lambda)^{-2n} },
\end{array}
\end{equation}
where
\begin{equation}
  \sum_{n=0}^{\infty}b_n(\kappa,N)\,z^{-2n} \sim
  e^{\displaystyle{N \sum_{m=1}^\infty a_m(\kappa)\,
  z^{-2m}}}.
\end{equation}
The coefficients $b_n(\kappa,N)$ are polynomials in $\kappa$ and
$N$ given by
\begin{equation}\label{bs-as}\begin{array}{c}
    \displaystyle{
      b_n(\kappa,N)=\sum_{r_1+2 r_2+\dots+n r_n=n}N^{r_1+ r_2+\dots+
    r_n} \, \times}\\ \\
    \displaystyle{ \times \,
    \frac{a_1(\kappa)^{r_1}\,a_2(\kappa)^{r_2}\,\dots\,
    a_n(\kappa)^{r_n}}
    {r_1!\, r_2!\, \dots \, r_n!} },
\end{array}
\end{equation}
where the sum extends over all sets of non negative integers
$r_1,r_2,\dots,r_n$ such that $r_1+2\, r_2+\dots+n\, r_n=n$. For
the  first five coefficients we get
\begin{equation}\label{prim-b}\begin{array}{c}
  b_0(\kappa,N)=1,\\ \\
  b_1(\kappa,N)=\frac{8 }{3}\,N\,\kappa \,\left( 1 - 3\,\kappa  +
      2\,{\kappa }^2 \right),\\ \\
  b_2(\kappa,N)=\frac{32}{45}\,N\,\kappa \,\left( 5\,N\,\kappa \,
       {\left( 1 - 3\,\kappa  + 2\,{\kappa }^2 \right) }^
        2 +\right. \\ \\
        \left. + 6\,\left( -1 + 10\,{\kappa }^2 -
         15\,{\kappa }^3 + 6\,{\kappa }^4 \right)
      \right) \\ \\
  b_3(\kappa,N)=\frac{256}{2835}\,
  N\, \kappa \, \left( 1 - 3\, \kappa  + 2\, {\kappa }^2 \right)
  \times\\ \\
  \times \left(
      360 - 18\, \left( -60 + 7\, N \right) \, \kappa  + 35\,
      N^2\, {\kappa }^2  -\right.\\ \\-\left.
        30\, \left( 72 - 42\, N + 7\, N^2 \right) \, {\kappa }^3
        +\right. \\ \\\left.+
        5\, \left( 216 - 378\, N + 91\, N^2 \right) \, {\kappa }^4
        -\right. \\ \\
        \left. - 84\,
      N\, \left( -9 + 5\, N \right) \, {\kappa }^5 + 140\,
      N^2\, {\kappa }^6 \right) , \\ \\
      b_4(\kappa,N)=\frac{512}{42525}\, N\, \kappa \, \left(
        1 - 3\, \kappa  + 2\, {\kappa }^2 \right) \times \\ \\
        \times \left( -45360 +
        36\, \left( -3780 + 221\, N \right) \, \kappa  - \right.
        \\ \\ \left. -
        252\, \left( 60 - 9\, N + 5\, N^2 \right) \, {\kappa }^2 +
        \right. \\ \\ \left. +
        7\, \left( 32400 - 8604\, N + 540\, N^2 + 25\,
            N^3 \right) \, {\kappa }^3 - \right. \\ \\
            \left. -
        315\, \left( -240 + 36\, N - 32\, N^2 + 5\,
            N^3 \right) \, {\kappa }^4 + \right. \\ \\
            \left. +
        21\, \left( -10800 + 9684\, N - 2700\, N^2 + 275\,
            N^3 \right) \, {\kappa }^5 - \right. \\ \\
            \left. -
        63\, \left( -1200 + 3156\, N - 1420\, N^2 + 175\,
            N^3 \right) \, {\kappa }^6 + \right. \\ \\
            \left. + 6\,
      N\, \left( 9468 - 10080\, N + 1925\, N^2 \right) \, {\kappa }^7
      - \right. \\ \\ \left. -
        1260\, N^2\, \left( -12 + 5\, N \right) \, {\kappa }^8 + 1400\,
      N^3\, {\kappa }^9 \right).
\end{array}
\end{equation}

Now, replacing eq.\ (\ref{GsGN}) in eq.\ (\ref{asymp}) we get
\begin{equation}\label{des-asymp-fin}\begin{array}{c}
  \displaystyle{
    \frac{\left[\psi\left(1-\kappa-\frac{\lambda}{4}\right)-
            \psi\left(\kappa-\frac{\lambda}{4}\right)\right]}
            {4\left[1-\beta\frac{\Gamma\left(1-\kappa-
            \frac{\lambda}{4}\right)}
            {\Gamma\left(\kappa-\frac{\lambda}{4}\right)}\right]}\sim
            - \sum_{N=1}^\infty \sum_{n=0}^\infty
  4^{N(2\kappa-1)}\,\times}
            \\ \\
  \displaystyle{ \times\, \beta^{N}\left( 2\kappa-1+\frac{2n}{N}
  \right)\, b_n(\kappa,N)\, \left( -\lambda \right)^{-N(2\kappa-1) - 2n
  -1}\equiv}\\ \\
  \displaystyle{\equiv \sum_{N=1}^\infty \sum_{n=0}^\infty C_{N,n}
  (\kappa,\beta)\,
  \left( -\lambda \right)^{-N(2\kappa-1) - 2n -1}}.
\end{array}
\end{equation}

Finally, eqs.\ (\ref{asymp-psi}) and (\ref{des-asymp-fin}) lead to
the asymptotic expansion for $f'(\lambda)/f(\lambda)$ in eq.\
(\ref{te}).

%%%%%%%%%%%%%%%%%%%%%%%%%%%%%%%%%%%%%%%%%

% ----------------------------------------------------------------

\begin{references}
\bibitem{foot}This potential
corresponds to a classically integrable system (see
\cite{Perelomov}, for example). The spectrum of the quantum model
subject to Dirichlet boundary conditions has been determined in
\cite{Calogero}. This Hamiltonian also appears as an effective
radial operator for an isotropic harmonic oscillator in
multi-dimensional Euclidean space with given angular momentum
eigenvalue \cite{Peak}, problem for which several results are
known: See \cite{Peak,Kleinert,Mueller} for the heat-kernel and
the resolvent in the case of Dirichlet boundary conditions. The
resolvent for a different boundary condition could be obtained
from this one by means of the so called Krein's formula
\cite{Achiezer,Albeverio2}. Our approach to the construction of
the different SAE of this symmetric Hamiltonian will rather be
based on von Neumann's theory \cite{Reed-Simon}, after the
explicit determination of the deficiency subspaces.
\end{references}
\end{document}